\documentclass[runningheads]{llncs}
\usepackage[T1]{fontenc}
\usepackage{graphicx}
\usepackage{amsmath,amssymb}
\usepackage{amsfonts}
\usepackage{xcolor}
\usepackage{comment}
\usepackage{multirow}
\usepackage{booktabs}
\usepackage{threeparttable}

\usepackage{enumitem}

\begin{document}
\title{Comparative Analysis of Formula and Structure Prediction from Tandem Mass Spectra}
\author{Xujun Che \and Xiuxia Du* \and Depeng Xu*}

\institute{University of North Carolina at Charlotte\\
\email{{\{xche,xdu4,dxu7\}@charlotte.edu}}}
\maketitle              
\begin{abstract}
Liquid chromatography mass spectrometry (LC-MS)-based metabolomics and exposomics aim to measure detectable small molecules in biological samples. The results facilitate hypothesis-generating discovery of metabolic changes and disease mechanisms and provide information about environmental exposures and their effects on human health. Metabolomics and exposomics are made possible by the high resolving power of LC and high mass measurement accuracy of MS. However, a majority of the signals from such studies still cannot be identified or annotated using conventional library searching because existing spectral libraries are far from covering the vast chemical space captured by LC-MS/MS. To address this challenge and unleash the full potential of metabolomics and exposomics, a number of computational approaches have been developed to predict compounds based on tandem mass spectra. Published assessment of these approaches used different datasets and evaluation. To select prediction workflows for practical applications and identify areas for further improvements, we have carried out a systematic evaluation of the state-of-the-art prediction algorithms. Specifically, the accuracy of formula prediction and structure prediction was evaluated for different types of adducts. The resulting findings have established realistic performance baselines, identified critical bottlenecks, and provided guidance to further improve compound predictions based on MS.

\keywords{untargeted metabolomics  \and mass spectrometry \and compound prediction \and deep learning.}
\end{abstract}

\section{Introduction}

Metabolomics and exposomics, the large-scale study of small molecules within cells, tissues, or organisms, is pivotal for understanding complex biological systems and discovering novel biomarkers. Liquid chromatography coupled to mass spectrometry is one of the major analytical platforms for metabolomics and exposomics studies due to its broad coverage of the chemical space. A fundamental task in mass spectrometry-based metabolomics and exposomics is to assign compounds or compound classes to tandem mass spectra, essential for translating raw analytical data into actionable biological insights. However, this is still a  challenge and a majority of the resulting tandem mass spectra remain unknown, forming the heavy and large dark matter that needs to be elucidated to fulfill the true potential of mass spectrometry-based metabolomics and exposomics.

LC separates compounds in complex mixtures based on their chemical properties. This separation reduces chemical complexity, prevents different analytes from interfering with each other, and enables more accurate detection of compounds. The subsequent tandem mass spectrometry (MS/MS) provides structural information of compounds by fragmenting corresponding precursor ions to produce characteristic fragmentation patterns that are the basis for molecular identification. Traditionally, an unknown MS/MS spectrum is matched against a library of known MS/MS. Compounds of the most similar MS/MS in the library are assigned to the unknown MS/MS. This process is known as compound identification (confident assignment with strong evidence) or annotation (assignment with some evidence). The library search-based method has proven to be very effective. However, existing MS/MS libraries is still tiny compared to the vast space of compounds that LC-MS platforms are capable of detecting. As a result, the majority of MS/MS in a typical metabolomics or exposomics study remain unknown after library search.

To address this challenge, computational workflows have been developed to predict compounds based on MS/MS. Compound prediction overcomes the limitation of MS/MS libraries and can provide invaluable insights about what compounds could have produced an unknown MS/MS, even though it cannot achieve the level of confidence that library matching could provide. Compound prediction typically consists of two sequential steps: \textit{molecular formula prediction} and \textit{structure elucidation}. 

For molecular formula prediction, a list of potential candidate formula are generated and ranked. One such algorithm is Sirius~\cite{duhrkop2019sirius}. It combines isotope pattern modeling with probabilistic fragmentation-tree construction from MS/MS spectra, scores all candidates to select the molecular formula most consistent with both MS1 and MS/MS evidence. It produces interpretable results, but struggles with speed and complexity. To address these limitations, deep learning-based methods such as MIST-CF \cite{goldman2023mist} and FIDDLE \cite{hong2024fiddle} have emerged as alternatives. These data-driven models have shown improved performance and reasonable accuracy by capturing complex spectral patterns, although they typically require large training datasets and can be less transparent in their predictions.

For each formula candidate, structure prediction can be carried out using two strategies: database searching or de novo generation. The former method is used by CSI:FingerID that predicts the molecular structure fingerprint based on the fragmentation tree that SIRIUS has generated and then computes the similarity between the predicted fingerprint and that of known compounds in a structural database \cite{duhrkop2015searching}. Understandably, this method is limited to known molecules and would miss novel structures—a critical limitation given that a large percentage of spectra in typical metabolomics experiments are believed to originate from unknown compounds not present in existing structural libraries \cite{da2015illuminating,monge2019challenges,bittremieux2022critical}. Methods that generate de novo structures include MADGEN \cite{wang2025madgen}, MSNovelist \cite{stravs2022msnovelist}, and DiffMS \cite{bohde2025diffms}, among others. These de novo prediction approaches struggle with the high complexity of the task, especially in distinguishing between structural isomers. Even with high-resolution mass spectrometry, multiple formula candidates can remain indistinguishable, a problem that is compounded when trying to identify isobaric or isomeric compounds \cite{bohde2025diffms}.

Beyond these intrinsic technical challenges, current methodologies for evaluating computational models often face limitations that question their practical utility. Firstly, many benchmark datasets suffer from issues of scale and coverage. For instance, the widely used Global Natural Products Social Molecular Networking (GNPS) database \cite{wang2016sharing} contains spectra for only an estimated 2.5\% of known natural products \cite{de2023ms2query,van2020linking}, suggesting that models trained and evaluated on such datasets may not be representative of the broader chemical universe. Secondly, a prevalent evaluation paradigm is the use of stringent structure-dissimilarity splits \cite{bushuiev2024massspecgym}, where molecules in the test set are intentionally chosen to be structurally distinct from those in the training set. While this approach rigorously tests a model's generalization capabilities, it overemphasizes this aspect at the expense of practical applicability. In many real-world metabolomics studies, researchers are focused on identifying metabolites within specific, known families where structural similarities are expected. The current focus on broad generalization does not reflect these common use-cases.

To bridge this gap between academic benchmarking and practical application, we evaluated the performance of state-of-the-art computational methods on large-scale, comprehensive datasets. Specifically, we conducted systematic evaluations using tandem mass spectra in the National Institute of Standards and Technology (NIST) 23  tandem spectral library \cite{nist} and the Mass Bank of North America (MoNA) \cite{mona}, representing both high-quality curated spectra and diverse community-contributed data. We implemented a two-stage pipeline that first predicts the molecular formula and then predicts the molecular structure for each of the proposed formula. We adopted random data splitting, rather than structure-dissimilarity splitting, to reflect practical metabolomics scenarios. Moreover, we performed adduct-type-stratified analysis to reveal performance heterogeneity across different adducts, providing actionable insights masked by aggregate statistics.

Our evaluation revealed that computational compound prediction has reached practically useful accuracy levels for dominant adduct types. For [M+H]$^+$, trained models achieved over 86\% formula prediction accuracy and over 67\% structure prediction accuracy. However, less common adduct types such as [M+H-2H$_2$O]$^+$ and [M-H+2Na]$^+$ exhibited severe performance degradation, highlighting fundamental challenges that require architectural innovations beyond simple data scaling. By evaluating structure generation under both oracle conditions where the ground truth formula are provided and realistic pipelines where formula are predicted, we were able to decompose the total prediction error into contributions from each stage. We found that errors from formula prediction propagate surprisingly modestly to structure generation. These findings provide a foundation for prioritizing future research directions and inform best practices for deploying these methods in real-world metabolomics applications.


\vspace{-6pt}
\section{Experimental Section}

\subsection{Datasets}

\subsubsection{\textbf{NIST 23 Spectral Library:}} A commercial, high-quality spectrometral library widely regarded as the gold standard in the field. It contains carefully curated spectra with rigorous quality control, making it an ideal benchmark dataset for evaluating prediction accuracy in controlled settings. We used its MS/MS spectra and filtered them based on the following six criteria:

\begin{enumerate}[label=(\arabic*), topsep=0pt]
    \item Ionization polarity: Positive.
    \item Fragmentation type: HCD.
    \item Charge state: Singly charged precursor ions only (charge = 1).
    \item Elements: H, C, N, O, P, S, F only (covering most biologically relevant metabolites).
    \item Adduct types: [M+H]$^+$, [M+Na]$^+$, [M+K]$^+$, [M+H-H$_2$O]$^+$, [M+H-2H$_2$O]$^+$, [M-H+2Na]$^+$.
    \item Collision energy: Up to three spectra per molecule-adduct pair, selecting those with median collision energies to reduce redundancy.
\end{enumerate}

\subsubsection{\textbf{MoNA (MassBank of North America:)}} An open-access mass spectrometry database that are contributed by the metabolomics community. Unlike NIST, MoNA represents the diversity and heterogeneity of real-world spectral data, including contributions from various laboratories with different experimental protocols and instrumentation. We downloaded the LC-MS/MS Positive Mode spectra and filtered them based on the following criteria: 

\begin{enumerate}[label=(\arabic*), topsep=0pt]
    \item Elements: H, C, N, O, P, S, F only.
    \item Adduct types: Same as NIST 23.
\end{enumerate}

\subsubsection{\textbf{Dataset Statistics:}}Table~\ref{tab:dataset_overview} summarizes the statistics of the resulting dataset and Table~\ref{tab:dataset_splits} shows the spectra distribution of adduct types in each dataset. The latter reveals significant imbalances that reflect real-world scenarios, where certain adducts dominate. Notably, [M+H]$^+$ constitutes the majority of spectra across both datasets, accounting for approximately 56.1\% in NIST, and 78.6\% in MoNA. In contrast, [M+H-2H$_2$O]$^+$ and [M-H+2Na]$^+$ represent only a small fraction of the data, with fewer than 30 spectra in some cases.

\begin{table}[h]
\centering
\caption{Overall statistics of the two datasets after filtering. (MW = Molecular Weight)}
\label{tab:dataset_overview}

\setlength{\tabcolsep}{4pt}
\scriptsize
\begin{tabular}{@{}lrr@{}}
\toprule
\textbf{Metric} & \textbf{NIST} & \textbf{MoNA} \\
\midrule
\# Spectra           & 86,435 & 76,196 \\
\# Unique Molecules  & 19,511 & 11,894 \\
\# Unique Formulas   &  9,630 &  6,244 \\
MW Range (Da)        & 82.11--1,704.01 & 45.08--2,680.17 \\
\bottomrule
\end{tabular}
\end{table}

\subsection{Computational Methods for Compound Prediction}


\subsubsection{Chemical Formula Prediction}
\label{sec:formula_methods}

\paragraph{\textbf{SIRIUS:}} The state-of-the-art fragmentation tree-based formula prediction \cite{duhrkop2019sirius}. It operates by first enumerating candidate chemical formulas consistent with the observed precursor mass within a specified mass tolerance. For each formula candidate, SIRIUS assigns potential subformulae to MS/MS peaks and constructs a fragmentation tree that represents plausible neutral loss pathways from the precursor to observed fragments. The fragmentation tree is scored using a maximum a posteriori (MAP) estimation framework that incorporates both fragment plausibility based on chemical knowledge and the quality of the MS/MS spectrum explanation. SIRIUS employs hand-crafted scoring functions parameterized with domain knowledge, including expected fragmentation patterns, common neutral losses, and isotopic distributions. The method can optionally incorporate MS1 isotope pattern information to improve formula discrimination, though we focus exclusively on the MS/MS-based scoring component for fair comparison with MIST-CF (to be described). 
It is worth noting that SIRIUS is a rule-based method with hand-crafted parameters and therefore does not require training.

\paragraph{\textbf{MIST-CF:}} A neural network-based approach for ranking chemical formula candidates given an MS/MS spectrum \cite{goldman2023mist}. It employs an energy-based modeling framework with a Chemical Formula Transformer architecture. Given a candidate chemical formula and its corresponding adduct type, MIST-CF first assigns plausible subformulae to MS/MS fragment peaks within a specified mass tolerance. These subformula-annotated peaks are then encoded using sinusoidal embeddings and processed through a transformer network that learns to score the compatibility between the candidate formula and the observed fragmentation pattern. Unlike fragmentation tree-based methods, MIST-CF learns scoring functions directly from data without requiring explicit tree construction, enabling efficient evaluation of large candidate sets. For our experiments, we utilized a model pretrained on the NPLIB1 dataset following the MIST-CF procedure with adjustments to focus on our six adduct types of interest. Additionally, we trained two separate models from scratch on NIST and MoNA datasets respectively.


\subsubsection{Molecular Structure Generation}
\label{sec:structure_methods}

\paragraph{\textbf{MSNovelist:}} An LSTM-based sequence-to-sequence model that generates molecular structures in SMILES string representation conditioned on molecular fingerprint predictions \cite{stravs2022msnovelist}. It operates in two stages: first, a fingerprint prediction model infers a probabilistic molecular fingerprint from the MS/MS spectrum; second, an LSTM decoder autoregressively generates SMILES characters guided by the predicted fingerprint. For our experiments, we used the MSNovelist implementation from the SIRIUS software. The underlying RNN model was trained on 1,232,184 chemical structures from HMDB, COCONUT, and DSSTox databases.

\paragraph{\textbf{DiffMS:}} A conditional molecular generation model based on discrete graph diffusion \cite{bohde2025diffms}. The method consists of two main components: a spectrum encoder and a graph decoder. The encoder uses the same Formula Transformer architecture as MIST-CF to extract structural information from MS/MS spectra, producing a fixed-dimensional embedding that captures the fragmentation pattern. The decoder implements a discrete diffusion process on molecular graphs, where bond types are iteratively denoised starting from a random initialization, conditioned on both the spectrum embedding and the known chemical formula.


For our experiments, we utilized the pretrained DiffMS model released by the authors, which was trained on NPLIB1 data. For domain adaptation, we trained two separate models initialized with the pretrained decoder weights on NIST and MoNA datasets respectively. We evaluated DiffMS under two experimental settings to isolate different sources of error in the complete prediction pipeline:


\begin{enumerate}[label=(\arabic*)]
    \item {\textbf{DiffMS (Oracle}):} In this configuration, we provided the ground truth chemical formula to the DiffMS decoder. This setting represents the upper bound of structure prediction performance, isolating the decoder's capability to generate correct molecular graphs given correct formula. Any errors in this setting reflect limitations in the structure generation model itself or ambiguities inherent in the MS/MS data (e.g., inability to distinguish stereoisomers or structural isomers with similar fragmentation patterns).
    
    \item \textbf{DiffMS (MIST-CF):} In this configuration, we constructed a complete end-to-end pipeline where MIST-CF first predicted candidate chemical formulas from the MS/MS spectrum, and the top-5 ranked formula predictions were then each fed to the DiffMS decoder for structure generation. For each of the five formula, DiffMS generated molecular structures, resulting in an expanded set of structure candidates that accounts for uncertainty in the formula prediction step. We ranked the final structure candidates by structure generation frequency. Performance in this setting reflects the complete annotation pipeline and quantifies how formula prediction uncertainty propagates through to structure generation. 
    
    

    \end{enumerate}

For the Oracle setting, we generated 100 molecular structure candidates per spectrum. For the MIST-CF configuration, we generated 20 structure candidates per formula for each of the top-5 predicted formulas, yielding a total of up to 100 candidates per spectrum.

\vspace{-6pt}
\subsection{Evaluation Setup}
\label{sec:eval_setup}

\subsubsection{Metrics:} We employed standard metrics from the mass spectrometry literature. For formula prediction, we computed the top-$K$ accuracy. For structure prediction, we used two metrics: (1) top-$K$ accuracy based on InChIKey-14 match (allowing stereoisomer variation), and (2) maximum Tanimoto similarity. All metrics report the best match within top-$K$ predictions.

\subsubsection{Data Split:} We split both NIST and MoNA datasets into training, validation, and test sets through random splitting. This strategy was deliberately chosen over structure-dissimilarity splitting as it preserves the natural distribution of adduct types and chemical properties, thereby reflecting realistic application scenarios where compounds of interest often share structural features. 
Both datasets exhibit severe class imbalance across adduct types, with [M+H]$^+$ dominating while several rare adduct types are significantly underrepresented. 
Table~\ref{tab:dataset_splits} presents the split statistics after applying this balancing strategy.

\vspace{-6pt}
\begin{table}
\centering
\caption{Dataset split statistics by adduct types. }
\label{tab:dataset_splits}
\setlength{\tabcolsep}{4pt}
\scriptsize
\begin{tabular}{@{}lrrrrrrrrr@{}}
\toprule
                     & \multicolumn{4}{c}{\textbf{NIST}} & & \multicolumn{4}{c}{\textbf{MoNA}} \\
\cmidrule(lr){2-5} \cmidrule(lr){7-10}
\textbf{Adduct Type} & \textbf{Total} & \textbf{Train} & \textbf{Val} & \textbf{Test} & & \textbf{Total} & \textbf{Train} & \textbf{Val} & \textbf{Test} \\
\midrule
{[}M+H{]}$^+$         & 50,182 & 40,073 & 1,143 & 100 & & 59,854 & 59,654 & 100 & 100 \\
{[}M+Na{]}$^+$        & 13,376 & 10,790 &   298 & 100 & & 10,997 & 10,797 & 100 & 100 \\
{[}M+K{]}$^+$         &    354 &    292 &     6 &  30 & &  5,197 &  4,997 & 100 & 100 \\
{[}M+H-H$_2$O{]}$^+$  & 17,433 & 13,912 &   441 & 100 & &    102 &     81 &  10 &  11 \\
{[}M+H-2H$_2$O{]}$^+$ &  4,133 &  3,295 &    91 & 100 & &     21 &     16 &   2 &   3 \\
{[}M-H+2Na{]}$^+$     &    957 &    786 &    21 &  91 & &     25 &     20 &   2 &   3 \\
\midrule
\textbf{Total}        & 86,435 & 69,148 & 2,000 & 521 & & 76,196 & 75,565 & 314 & 317 \\
\bottomrule
\end{tabular}
\end{table}
\vspace{-16pt}

\section{Results and Discussion}
\label{sec:results}

\subsection{Formula Prediction: MIST-CF vs SIRIUS}
\label{sec:formula_results}

\subsubsection{Overall Performance Comparison:}
\label{sec:formula_overall}

Table~\ref{tab:formula_overall} summarizes the formula prediction accuracy of SIRIUS and MIST-CF on the NIST and MoNA datasets. We report top-1, top-5, and top-10 accuracy metrics, reflecting the practical scenario where domain experts typically inspect multiple top-ranked candidates. MIST-CF is evaluated in two configurations: \textit{pretrained} on external datasets (NPLIB1), and \textit{retrained} (in-domain) on the target dataset's training split. SIRIUS achieves strong performance on NIST (top-5: 0.758) but limited performance on MoNA (0.435). Pretrained MIST-CF shows moderate performance (0.655--0.685 top-5). In-domain retraining substantially improves MIST-CF: top-1 accuracy increases by 8.2 points (NIST) and 34.2 points (MoNA), with MoNA reaching 0.892 top-5 accuracy. However, NIST retraining shows an unexpected pattern: while top-1 improves (0.378 vs 0.296), top-5 and top-10 decline (0.503 vs 0.685 and 0.622 vs 0.727). Three factors may explain this: (1) MIST-CF's contrastive training prioritizes top-1 separation over ranking calibration; (2) NIST's controlled, high-quality spectra enable learning of dataset-specific artifacts that improve top-1 but reduce ranking robustness; (3) smaller NIST improvements versus MoNA (8.2\% vs 34.2\%) suggest NPLIB1 already contains similar data, limiting retraining benefits while introducing overfitting.

\begin{table}
\centering
\caption{Overall formula prediction accuracy of SIRIUS and MIST-CF. SIRIUS does not require training, while MIST-CF is evaluated for both its base performance and domain-adaptation retraining performance. Bold values indicate the best performance for each testing configuration.}
\label{tab:formula_overall}

\scriptsize
\begin{tabular}{@{}lccccccc@{}}
\toprule
                        & \multicolumn{3}{c}{\textbf{NIST}} & & \multicolumn{3}{c}{\textbf{MoNA}} \\
\cmidrule(lr){2-4} \cmidrule(lr){6-8}
\textbf{Method}         & Top-1  & Top-5  & Top-10 & & Top-1  & Top-5  & Top-10 \\
\midrule
SIRIUS                  & 0.284  & \textbf{0.758} & \textbf{0.825} & & 0.193 & 0.435 & 0.500 \\
MIST-CF (pretrained)    & 0.296  & 0.685  & 0.727  & & 0.402 & 0.655 & 0.744 \\
MIST-CF (retrained)     & \textbf{0.378} & 0.503 & 0.622 & & \textbf{0.744} & \textbf{0.892} & \textbf{0.915} \\
\bottomrule
\end{tabular}
\end{table}

\vspace{-10pt}
\subsubsection{Adduct-Type-Stratified Formula Prediction Performance:}
\label{sec:formula_adduct}

While aggregate metrics provide a useful overview, they obscure critical performance variations across adduct types. Table~\ref{tab:formula_adduct_stratified} presents top-5 formula prediction accuracy broken down by adduct types across all evaluation configurations, revealing substantial disparities that have important practical implications. 


\begin{table}[t]
\centering
\caption{Top-5 formula prediction accuracy stratified by adduct type.}
\label{tab:formula_adduct_stratified}
\scriptsize
\begin{tabular*}{\textwidth}{@{}l@{\extracolsep{\fill}}cccccc@{}}
\toprule
                        & \multicolumn{3}{c}{\textbf{NIST}} & \multicolumn{3}{c}{\textbf{MoNA}} \\
\cmidrule(lr){2-4} \cmidrule(lr){5-7}
\textbf{Adduct Type}    & SIRIUS & \begin{tabular}[c]{@{}c@{}}MIST-CF\\(pretrained)\end{tabular} & \begin{tabular}[c]{@{}c@{}}MIST-CF\\(retrained)\end{tabular} & SIRIUS & \begin{tabular}[c]{@{}c@{}}MIST-CF\\(pretrained)\end{tabular} & \begin{tabular}[c]{@{}c@{}}MIST-CF\\(retrained)\end{tabular} \\
\midrule
{[}M+H{]}$^+$               & 0.940 & 0.970 & \textbf{0.990} & 0.730 & \textbf{0.869} & \textbf{0.869} \\
{[}M+Na{]}$^+$              & 0.750 & 0.840 & \textbf{0.950} & 0.520 & 0.780 & \textbf{0.930} \\
{[}M+K{]}$^+$               & 0.667 & 0.233 & \textbf{0.800} & 0.160 & 0.310 & \textbf{0.990} \\
{[}M+H-H$_2$O{]}$^+$        & \textbf{0.950} & 0.850 & 0.160 & 0.727 & \textbf{0.909} & 0.364 \\
{[}M+H-2H$_2$O{]}$^+$       & \textbf{0.950} & 0.840 & 0.150 & \textbf{1.000} & 0.667 & 0.000 \\
{[}M-H+2Na{]}$^+$           & \textbf{0.176} & 0.000 & 0.143 & 0.000 & 0.000 & 0.000 \\
\bottomrule
\end{tabular*}
\end{table}

\noindent\textit{Dominant Adducts Achieve High Performance:} For [M+H]$^+$ and [M+Na]$^+$, retrained MIST-CF achieves >0.86 accuracy on both datasets, with NIST reaching 0.990 and 0.950 respectively. SIRIUS also performs well on these common types (0.730--0.940), confirming that formula annotation has reached practical utility for the majority of metabolomics applications.

\vspace{+5pt}
\noindent\textit{Water-Loss Adducts Show Unexpected Patterns:} Remarkably, water-loss adducts ([M+H-H$_2$O]$^+$ and [M+H-2H$_2$O]$^+$) exhibit reversed performance trends: SIRIUS achieves the highest accuracy (0.727--1.000), while retrained MIST-CF shows catastrophic failure (0.000--0.364). On NIST, SIRIUS reaches 0.950 for both types, whereas retrained MIST-CF collapses to 0.160 and 0.150—a dramatic 80-percentage point decline from pretrained performance (0.850 and 0.840). This suggests that: (1) rule-based fragmentation tree methods are better suited for handling water-loss ambiguities, where multiple formulas can produce similar mass shifts; (2) NIST retraining may have overfitted to dominant adduct patterns, losing the ability to generalize to chemically ambiguous cases; (3) small training sample sizes for these types (Table~\ref{tab:dataset_splits}) exacerbate overfitting.

\vspace{+5pt}
\noindent\textit{Extreme Rarity: [M-H+2Na]$^+$ Remains Unsolved:} The [M-H+2Na]$^+$ adduct shows near-zero accuracy across all methods (0.000--0.176), reflecting the extreme data scarcity and also raising the question about whether or not the adduct type has been determined correctly.

\vspace{-10pt}
\subsection{Structure Prediction: DiffMS vs MSNovelist}
\label{sec:structure_results}

\subsubsection{Overall Performance:}
\label{sec:structure_overall}

Table~\ref{tab:structure_overall} presents molecular structure prediction performance across different configurations.

\begin{table}[t]
\centering
\caption{Overall structure prediction performance (Top-5 metrics). DiffMS (Oracle) uses ground truth formulas while DiffMS (MIST-CF) uses MIST-CF's top-5 formula predictions. MSNovelist is pretrained and independent of our training configurations.}
\label{tab:structure_overall}

\setlength{\tabcolsep}{4pt}
\scriptsize
\begin{tabular}{@{}lcccccc@{}}
\toprule
                        & \multicolumn{2}{c}{\textbf{NIST}} & & \multicolumn{2}{c}{\textbf{MoNA}} \\
\cmidrule(lr){2-3} \cmidrule(lr){5-6}
\textbf{Method}         & Accuracy & Tanimoto & & Accuracy & Tanimoto \\
\midrule
MSNovelist                          & 0.447    & 0.619    &  & 0.249    & 0.405    \\
DiffMS (pretrained, Oracle)         & 0.077    & 0.310    &  & 0.088    & 0.298    \\
DiffMS (pretrained, MIST-CF)        & 0.052    & 0.282    &  & 0.066    & 0.272    \\
DiffMS (retrained, Oracle)          & \textbf{0.664} & \textbf{0.818} &  & \textbf{0.634} & 0.812    \\
DiffMS (retrained, MIST-CF)         & 0.662    & 0.817    &  & 0.630    & \textbf{0.818} \\
\bottomrule
\end{tabular}
\end{table}

\noindent\textit{Error Propagation Analysis:} A striking finding is that the retrained DiffMS (MIST-CF) pipeline performs remarkably close to the DiffMS (Oracle) setting. Accuracy gaps between the Oracle and MIST-CF configurations are minimal: 0.2 percentage points on NIST (0.664 vs 0.662) and 0.4 points on MoNA (0.634 vs 0.630). This robustness is surprising given that formula prediction is imperfect (e.g., MIST-CF retrained Top-5 accuracy on NIST is 0.503, see Table~\ref{tab:formula_overall}). This resilience is likely due to two factors: (1) the multi-formula strategy (generating structures for the Top-5 formulas) mitigates errors, and (2) the spectrum embedding provides structural constraints that partially compensate for incorrect formulas. This suggests that structure generation, not formula prediction (within the top-5), is the primary bottleneck. The Tanimoto scores are virtually unchanged, confirming that even when the pipeline fails, the generated structures remain chemically relevant.

\begin{figure}[t]
\centering
\includegraphics[width=\textwidth]{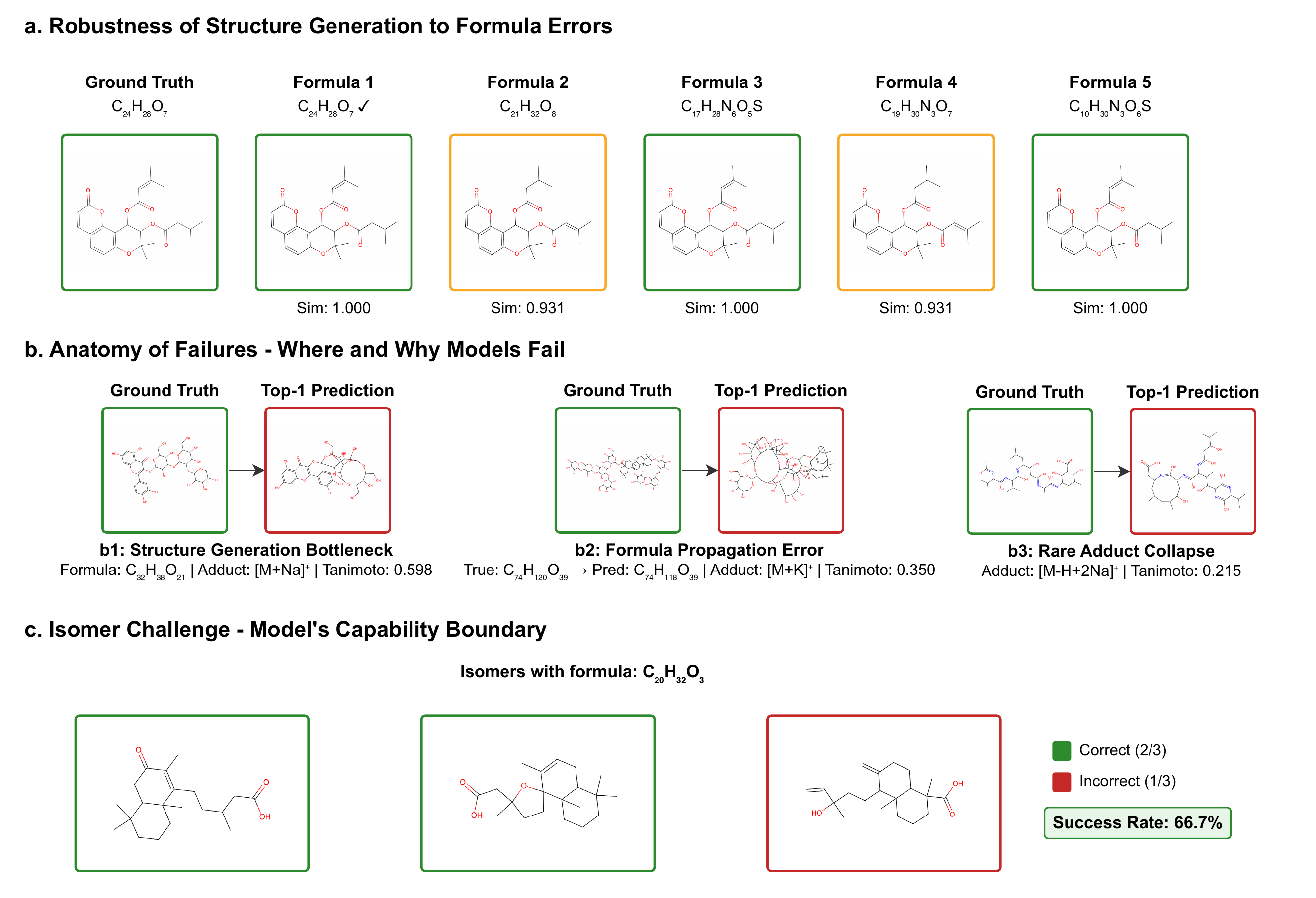}
\caption{Diagnostic analysis of pipeline performance. (a) Structure generation robustness to formula errors. (b) Failure mode decomposition: structure generation bottleneck (b1), formula propagation error (b2), and rare adduct collapse (b3). (c) Isomer discrimination challenge with 66.7\% success rate for C$_{20}$H$_{32}$O$_3$.}
\label{fig:error_analysis}
\end{figure}

To further understand the pipeline's error characteristics, we conducted a detailed failure analysis (Figure~\ref{fig:error_analysis}). Figure~\ref{fig:error_analysis}a demonstrates that structure generation is robust to formula errors, maintaining high Tanimoto similarity even for incorrect formulas. Figure~\ref{fig:error_analysis}b decomposes failure modes into structure generation bottlenecks (b1), formula propagation errors (b2), and rare adduct collapse (b3). Figure~\ref{fig:error_analysis}c shows that isomer discrimination remains challenging, with only 66.7\% accuracy for C$_{20}$H$_{32}$O$_3$ isomers.

\vspace{+5pt}
\noindent\textit{Domain Adaptation Impact:} Domain adaptation is critical for structure prediction. In-domain (retrained) models vastly outperform the general (pretrained) models. Comparing the oracle configurations, the accuracy gap between the retrained and pretrained DiffMS is 58.7 percentage points on NIST (0.664 vs 0.077) and 54.6 points on MoNA (0.634 vs 0.088). This sensitivity to domain adaptation is significantly larger than that observed for formula prediction, reflecting the greater complexity of structure generation.

\vspace{-10pt}
\subsubsection{Adduct-Type-Stratified Structure Prediction Performance:}
\label{sec:structure_adduct}

Table~\ref{tab:structure_adduct_stratified} \\presents top-5 accuracy stratified by adduct type.

\begin{table}[th]
\centering
\begin{threeparttable}
\caption{Top-5 structure prediction accuracy stratified by adduct type.}
\label{tab:structure_adduct_stratified}
\scriptsize
\begin{tabular*}{\textwidth}{@{}l@{\extracolsep{\fill}}lccccc@{}}
\toprule
                 &                      &                      & \multicolumn{4}{c}{\textbf{DiffMS}} \\
\cmidrule(lr){4-7}
\textbf{Dataset} & \textbf{Adduct Type} & \textbf{MSNovelist} & pretrained\textsuperscript{†} & pretrained\textsuperscript{‡} & retrained\textsuperscript{†} & retrained\textsuperscript{‡} \\
\midrule
\multirow{6}{*}{NIST} & {[M+H]$^+$           } & \textbf{0.800} & 0.170 & 0.170 & 0.730 & 0.760 \\
                      & {[M+Na]$^+$          } & 0.410 & 0.050 & 0.040 & 0.530 & \textbf{0.540} \\
                      & {[M+K]$^+$           } & 0.300 & 0.067 & 0.033 & \textbf{0.667} & \textbf{0.667} \\
                      & {[M+H-H$_2$O]$^+$    } & 0.550 & 0.040 & 0.010 & \textbf{0.750} & 0.720 \\
                      & {[M+H-2H$_2$O]$^+$   } & 0.320 & 0.010 & 0.000 & 0.560 & \textbf{0.570} \\
                      & {[M-H+2Na]$^+$       } & 0.176 & 0.132 & 0.055 & \textbf{0.758} & 0.725 \\
\midrule
\multirow{6}{*}{MoNA} & {[M+H]$^+$           } & 0.470 & 0.270 & 0.202 & \textbf{0.680} & 0.677 \\
                      & {[M+Na]$^+$          } & 0.150 & 0.010 & 0.020 & \textbf{0.620} & \textbf{0.620} \\
                      & {[M+K]$^+$           } & 0.090 & 0.000 & 0.000 & \textbf{0.610} & 0.600 \\
                      & {[M+H-H$_2$O]$^+$    } & 0.455 & 0.000 & 0.000 & \textbf{0.818} & \textbf{0.818} \\
                      & {[M+H-2H$_2$O]$^+$   } & \textbf{1.000} & 0.000 & 0.000 & 0.333 & 0.333 \\
                      & {[}M-H+2Na{]}$^+$     & 0.000 & 0.000 & 0.000 & 0.000 & 0.000 \\
\bottomrule
\end{tabular*}

\begin{tablenotes}
\small
\item[†] Oracle: structure prediction with ground truth formula;
\item[‡] Pipeline: with MIST-CF predicted formulas.
\end{tablenotes}

\end{threeparttable}
\end{table}

\noindent\textit{Dominant Adduct Types Achieve Practical Accuracy:} For [M+H]$^+$, the dominant adduct type, DiffMS (MIST-CF) achieves over 0.67 top-5 accuracy in in-domain settings on both datasets. Combined with >0.86 formula prediction accuracy, this demonstrates practically useful performance for the most common scenarios.

\noindent\textit{Severe Degradation for Rare Adduct Types:} The rare adduct types, [M+H-2H$_2$O]$^+$ and [M-H+2Na]$^+$, achieved only 0.333 and 0.000 top-5 accuracy on MoNA. 

\vspace{-6pt}
\subsection{Data Augmentation for Rare Adduct Types}
\label{sec:augmentation}

Given the severe performance degradation for less common adduct types, we tested whether data scarcity was the limiting factor by supplementing MoNA's limited training examples with NIST spectra, with 2,000 {[}M+H-H$_2$O{]}$^+$, 4,032 {[}M+H-2H$_2$O{]}$^+$ and 951 {[}M-H+2Na{]}$^+$. Despite substantial increases in training data, augmentation failed to improve performance for any of the less common adduct types for both formula prediction and structure prediction. These results suggest that data quality and relevance are more critical than raw quantity.

\vspace{-6pt}
\subsection{Practical Implications and Recommendations}
\label{sec:practical}

\vspace{-6pt}
\subsubsection{Rethinking Evaluation Paradigms:}
Current benchmarking efforts emphasize structure-dissimilarity splits, prioritizing generalization over practical relevance. In real metabolomics, researchers often study related compound families where structural similarity is expected. Our random-split strategy better reflects such contexts and shows that with sufficient in-domain data, state-of-the-art models reach practical performance ($>0.86$ top-5 formula, $>0.67$ top-5 structure accuracy for [M+H]$^+$). We recommend reporting both in-domain and out-of-domain results to guide real-world applicability.

\vspace{-6pt}
\subsubsection{Data Curation Recommendations:} Augmentation experiments revealed three challenges: (1)Distributional mismatch between high-quality NIST spectra and heterogeneous MoNA data, limiting transferability; 
(2) Spectral ambiguity for water-loss adducts, making identification ill-posed; and (3) Limited model capacity for learning both dataset- and adduct-specific rules.Accordingly, we suggest:(1) Focus on in-domain data for the target chemical space; (2) Balance quality and diversity by mixing reference and community spectra; (3) Mitigate adduct imbalance via targeted acquisition for rare ionization modes; and (4)Adopt advanced augmentation (e.g., generative or domain-adaptive methods).

\vspace{-6pt}
\subsubsection{Future Research Directions}

Our work has identified several promising research directions: (1) Persistent gaps for less common adducts indicate a need for specialized architectures or meta-learning; (2) Integrating retention time or collision cross-section (CCS) could resolve ambiguities; and (3) Domain adaptation is essential for cross-source generalization.

\vspace{-6pt}
\section{Conclusion}
This study has demonstrated that compound prediction is now practical for routine use, though key bottlenecks remain. Adduct-stratified results have revealed that structure generation, not formula prediction, limits accuracy. Naïve data scaling fails for less common adducts, underscoring the need for specialized models, multimodal integration, and domain adaptation. Our analysis provides both validation for current tools and a roadmap toward more comprehensive, realistic compound prediction.

\bibliographystyle{splncs04}
\bibliography{reference}

@article{duhrkop2019sirius,
  title={SIRIUS 4: a rapid tool for turning tandem mass spectra into metabolite structure information},
  author={D{\"u}hrkop, Kai and Fleischauer, Markus and Ludwig, Marcus and Aksenov, Alexander A and Melnik, Alexey V and Meusel, Marvin and Dorrestein, Pieter C and Rousu, Juho and B{\"o}cker, Sebastian},
  journal={Nature methods},
  volume={16},
  number={4},
  pages={299--302},
  year={2019},
  publisher={Nature Publishing Group US New York}
}

@article{goldman2023mist,
  title={MIST-CF: chemical formula inference from tandem mass spectra},
  author={Goldman, Samuel and Xin, Jiayi and Provenzano, Joules and Coley, Connor W},
  journal={Journal of Chemical Information and Modeling},
  volume={64},
  number={7},
  pages={2421--2431},
  year={2023},
  publisher={ACS Publications}
}

@article{duhrkop2015searching,
  title={Searching molecular structure databases with tandem mass spectra using CSI: FingerID},
  author={D{\"u}hrkop, Kai and Shen, Huibin and Meusel, Marvin and Rousu, Juho and B{\"o}cker, Sebastian},
  journal={Proceedings of the National Academy of Sciences},
  volume={112},
  number={41},
  pages={12580--12585},
  year={2015},
  publisher={National Academy of Sciences}
}

@article{stravs2022msnovelist,
  title={MSNovelist: de novo structure generation from mass spectra},
  author={Stravs, Michael A and D{\"u}hrkop, Kai and B{\"o}cker, Sebastian and Zamboni, Nicola},
  journal={Nature Methods},
  volume={19},
  number={7},
  pages={865--870},
  year={2022},
  publisher={Nature Publishing Group US New York}
}

@article{bohde2025diffms,
  title={Diffms: Diffusion generation of molecules conditioned on mass spectra},
  author={Bohde, Montgomery and Manjrekar, Mrunali and Wang, Runzhong and Ji, Shuiwang and Coley, Connor W},
  journal={arXiv preprint arXiv:2502.09571},
  year={2025}
}

@article{de2023ms2query,
  title={MS2Query: reliable and scalable MS2 mass spectra-based analogue search},
  author={de Jonge, Niek F and Louwen, Joris JR and Chekmeneva, Elena and Camuzeaux, Stephane and Vermeir, Femke J and Jansen, Robert S and Huber, Florian and van der Hooft, Justin JJ},
  journal={Nature Communications},
  volume={14},
  number={1},
  pages={1752},
  year={2023},
  publisher={Nature Publishing Group UK London}
}

@article{van2020linking,
  title={Linking genomics and metabolomics to chart specialized metabolic diversity},
  author={van Der Hooft, Justin JJ and Mohimani, Hosein and Bauermeister, Anelize and Dorrestein, Pieter C and Duncan, Katherine R and Medema, Marnix H},
  journal={Chemical Society Reviews},
  volume={49},
  number={11},
  pages={3297--3314},
  year={2020},
  publisher={Royal Society of Chemistry}
}

@article{wang2016sharing,
  title={Sharing and community curation of mass spectrometry data with Global Natural Products Social Molecular Networking},
  author={Wang, Mingxun and Carver, Jeremy J and Phelan, Vanessa V and Sanchez, Laura M and Garg, Neha and Peng, Yao and Nguyen, Don Duy and Watrous, Jeramie and Kapono, Clifford A and Luzzatto-Knaan, Tal and others},
  journal={Nature biotechnology},
  volume={34},
  number={8},
  pages={828--837},
  year={2016},
  publisher={Nature Publishing Group US New York}
}

@online{mona,
  author       = {{Fiehn Laboratory}},
  title        = {MassBank of North America (MoNA)},
  url          = {https://mona.fiehnlab.ucdavis.edu/},
  urldate      = {2025-09-21},
  organization = {UC Davis Genome Center},
  year         = {2025}
}

@database{nist,
  author       = {{National Institute of Standards and Technology}},
  title        = {Tandem (MS/MS) Mass Spectral Library},
  version      = {NIST 23},
  publisher    = {Standard Reference Data Program},
  organization = {National Institute of Standards and Technology (NIST)},
  address      = {Gaithersburg, MD},
  year         = {2023},
  url          = {https://www.nist.gov/srd/nist-standard-reference-database-1a}
}

@article{da2015illuminating,
  title={Illuminating the dark matter in metabolomics},
  author={da Silva, Ricardo R and Dorrestein, Pieter C and Quinn, Robert A},
  journal={Proceedings of the National Academy of Sciences},
  volume={112},
  number={41},
  pages={12549--12550},
  year={2015},
  publisher={National Academy of Sciences}
}

@article{monge2019challenges,
  title={Challenges in identifying the dark molecules of life},
  author={Monge, Mar{\'\i}a Eugenia and Dodds, James N and Baker, Erin S and Edison, Arthur S and Fern{\'a}ndez, Facundo M},
  journal={Annual Review of Analytical Chemistry},
  volume={12},
  number={1},
  pages={177--199},
  year={2019},
  publisher={Annual Reviews}
}

@article{bittremieux2022critical,
  title={The critical role that spectral libraries play in capturing the metabolomics community knowledge},
  author={Bittremieux, Wout and Wang, Mingxun and Dorrestein, Pieter C},
  journal={Metabolomics},
  volume={18},
  number={12},
  pages={94},
  year={2022},
  publisher={Springer}
}

@article{hong2024fiddle,
  title={FIDDLE: a deep learning method for chemical formulas prediction from tandem mass spectra},
  author={Hong, Yuhui and Li, Sujun and Ye, Yuzhen and Tang, Haixu},
  journal={bioRxiv},
  pages={2024--11},
  year={2024},
  publisher={Cold Spring Harbor Laboratory}
}

@article{wang2025madgen,
  title={MADGEN: Mass-Spec attends to De Novo Molecular generation},
  author={Wang, Yinkai and Chen, Xiaohui and Liu, Liping and Hassoun, Soha},
  journal={arXiv preprint arXiv:2501.01950},
  year={2025}
}

@article{bushuiev2024massspecgym,
  title={MassSpecGym: A benchmark for the discovery and identification of molecules},
  author={Bushuiev, Roman and Bushuiev, Anton and de Jonge, Niek and Young, Adamo and Kretschmer, Fleming and Samusevich, Raman and Heirman, Janne and Wang, Fei and Zhang, Luke and D{\"u}hrkop, Kai and others},
  journal={Advances in Neural Information Processing Systems},
  volume={37},
  pages={110010--110027},
  year={2024}
}

\end{document}